# Using Maps and Machine Learning to Predict Economic Activity[1]

Imryoung Jeong[2]    Hyunjoo Yang[3]

April 1, 2022

## Abstract

We introduce a novel machine learning approach to leverage historical and contemporary maps and systematically predict economic statistics. Our simple algorithm extracts meaningful features from the maps based on their color compositions for predictions. We apply our method to grid-level population levels in Sub-Saharan Africa in the 1950s and South Korea in 1930, 1970, and 2015. Our results show that maps can reliably predict population density in the mid-20th century Sub-Saharan Africa using 9,886 map grids (5km by 5 km). Similarly, contemporary South Korean maps can generate robust predictions on income, consumption, employment, population density, and electric consumption. In addition, our method is capable of predicting historical South Korean population growth over a century.

[1] We thank Sangyoon Park and Adam Storeygard for generously sharing historical data of South Korea and African cities. Hyunjoo Yang acknowledges research support by National Research Foundation of Korea Research Grant (Grant number: #2021069641). All errors are ours.

[2] Sogang University, School of Economics, Baekbeom-ro 35, Mapo-gu, Seoul, South Korea (email address: imryoung@sogang.ac.kr).
[3] Sogang University, School of Economics, Baekbeom-ro 35, Mapo-gu, Seoul, South Korea (email address: hyang@sogang.ac.kr).

# 1. Introduction

For hundreds of years, maps have been used to effectively convey complex geographical and spatial information through abstract symbols, shapes, and text. Similar to contemporary remote sensing data, maps can provide consistent measurements across country borders. For example, the U.S. Army produced a series of 1:250,000 maps of regions covering most continents between the 1940s and the 1960s, now fully disclosed to the public (Figure 1).

Many important economic questions on growth and development hinge on accurate measurements of economic variables over the long run. Although remote sensing data, such as nighttime luminosity and day-time satellite images, have been used as reliable proxies for local economic activity, they have only become available in recent years, thus limiting their applicability for long-term analysis.

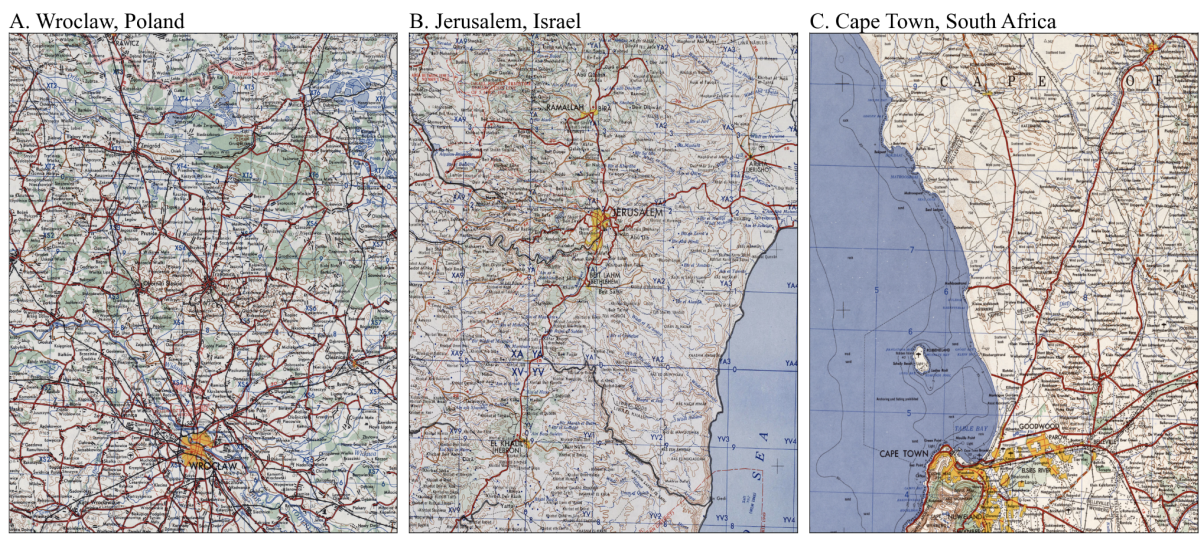

Figure 1: U.S. Army maps from different continents in the mid-20th century

In this paper, we introduce a novel solution of using historical and contemporary maps for predicting economic statistics. Our algorithm utilizes the distribution of colors used in full-color topographic maps. Unlike many machine learning methods, ours is computationally lightweight and easily replicable by social scientists. We show that both current and historical maps can be used to systematically predict various economic variables, including population density, income, consumption, and production. The predictive accuracy of our approach is comparable to that of convolutional neural network models, a conventional machine learning model for image classifications, in predicting almost every economic variable that we have used.

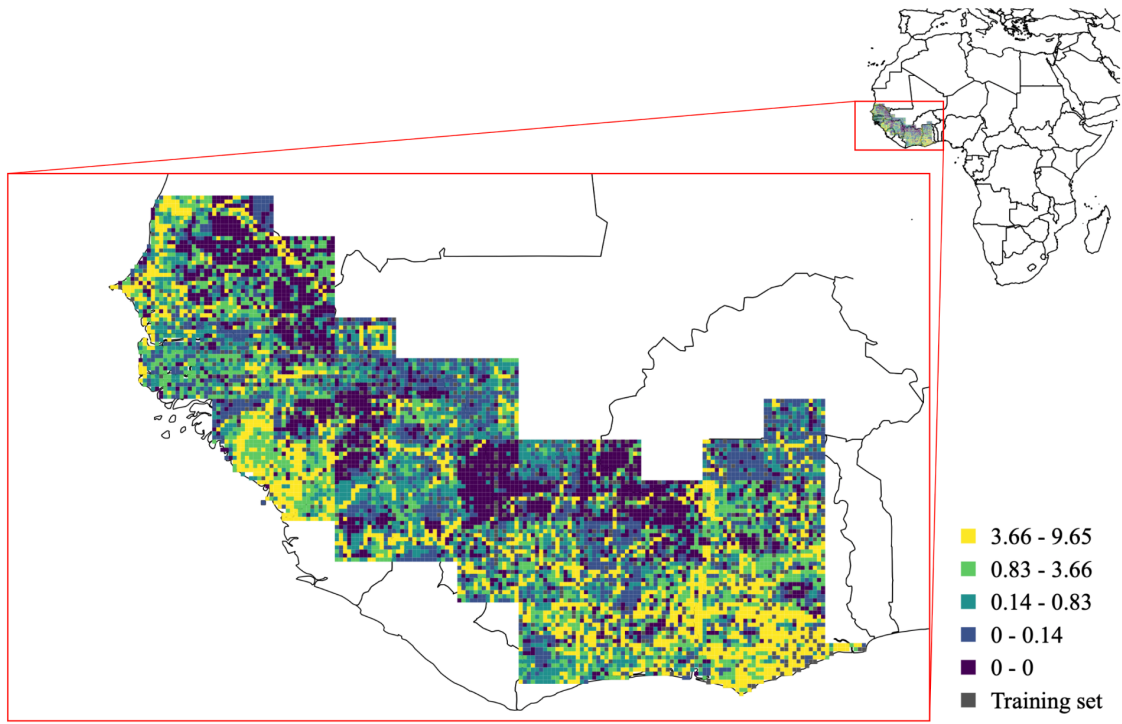

Figure 2. Predicted population density of selected African regions in 1950s using map color counts

To test our approach, we digitize and georeference current and historical maps of most parts of South Korea from two periods a century apart. In addition, we georeference 75 maps in Sub-Saharan Africa made by the American Army in the 1950s, and predict historical African population density reliably (Figure 2). When we check against high-quality grid-level and regional-level economic data from reliable sources produced by the South Korean government, our algorithm predicts income, consumption, employment, population density, and electric consumption with high accuracy after using only a small training set of approximately 900 images (20% of the total grids). In addition, our method is capable of predicting historical South Korean population growth over a century.

To the best of our knowledge, we are the first to use historical maps to systematically predict economic variables. Our research is closely related to growing literature in which scholars have used machine learning to recover historical datasets. For example, Feigenbaum, 2016 used classification algorithms to link the historical U.S. census data between 1915 and 1940, and Shen et al. (2020) applied deep learning-based layout parsers to extract information from historical Japanese documents. We make an important contribution to the literature by providing a general algorithm that can be applied to any georeferenced historical and contemporary maps to generate economic proxies. Despite their appeal, historical archives are underutilized due to the costly and labor-intensive nature of pre-processing data for analysis (see Combes et al., 2021 for the summary). We overcome this limitation by introducing a simple but effective approach that enables researchers to analyze data from full-color maps, both contemporary and historical.

More broadly, our study is related to literature in which researchers have used geospatial data to predict local economic measures such as income and assets (Jean et al., 2016) or urban areas (Baragwanath et al., 2021 ; Galdo et al., 2021). Some researchers have utilized Google Street View to measure the urban appearance and changes over time (Naik et al., 2016 ; Naik et al., 2017). Glaeser et al. (2018) showed that urban appearance can explain a considerable portion of the income levels in metropolitan areas in the United States. We contribute to this strand of literature by highlighting maps as another promising source of geospatial data that can be used to predict local economic activity.

## 2. Prediction Methods and Data

2.1 Methods

Recently, researchers have measured and predicted economic variables using geospatial datasets collected from mobile phone networks (Blumenstock et al., 2015) and digital platforms (Glaeser et al., 2018), or maps of buildings (Arribas-Bel et al., 2021 ; de Bellefon et al., 2021). Henderson et al. (2012) pioneered what has become one of the most popular approaches: utilizing remotely sensed luminosity at night to measure economic growth. While nighttime luminosity has shown promising improvements in measuring economic activities (Pinkovskiy and Sala-i-Martin, 2016), it appears less capable in less developed countries where the provision of electricity is lacking (Jean et al., 2016).

In a growing body of research, scholars have attempted to address this challenge by combining daytime satellite imagery with machine learning approaches (Jean et al., 2016 ; Yeh et al., 2020). However, the implementation of state-of-the-art machine learning models is a significant hurdle social scientists must overcome to leverage satellite images. Moreover, nighttime luminosity and daytime satellite images hardly fill the data gap between the recent and distant past. For example, nightlight luminosity data only date back to 1992, and high-resolution satellite images covering the entire globe, including Landsat, Sentinel, and Google Static Maps only date back to the mid-2010s.

Our approach using topographic maps involves three main steps. First, we convert the colors on the maps into a simple, standardized color palette, thereby enabling us to compare across the maps. Then, we count the number of pixels associated with each standardized color on a map to quantify its color composition. Finally, we train a linear regression model and a deep neural network model to predict the log population using the color counts from the previous step as inputs.

**Pre-processing.** We clip the georeferenced topographic maps into 5 km by 5 km grid cells to match the population grids from the National Geographic Information Institute (NGII) in South Korea.[4] Then, we merge the population density data with the grid cells. During the clipping and merging process, we remove grid cells located at the original maps' borders because they contain redundant areas (e.g., descriptions), which might disturb efficient learning. We also exclude grid cells with distorted areas caused by the reprojection process of the full-sized maps. The maps are georeferenced based on WGS84 (EPSG:4326), whereas the population grids are based on the Korea 2000/Unified CS (EPSG:5179) coordinate system, so distortion due to the reprojection process is unavoidable. As a result, the final dataset used in the analysis contained 4,452 grid cells.

---

[4] We provide a more detailed description on the data subsection below.

Step 1. Simplification
Original map
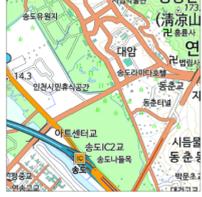
Total 222,090 colors

Simplified map (*k* = 10)
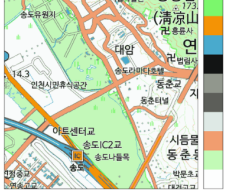

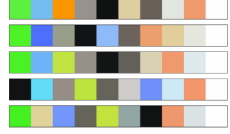

$n = 22,209$

$k = 10$

Step 2. Standardization
Standardized Color Palette (*k*=10)
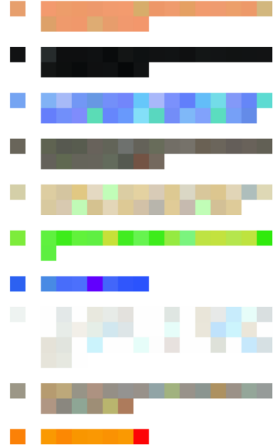

Standardized map
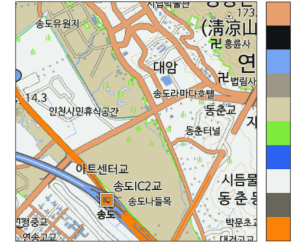

Figure 3: Our model's approach of reducing feature dimensions in maps

**Simplification.** We simplify the color palette through the simple K-means clustering. K-means clustering algorithm classifies the data points into a fixed number of clusters (*k*) without providing any guidance on how to classify them. Starting with random choices of the center of clusters (i.e., *centroids*), the algorithm allocates anonymous data points to the closest cluster based on the distance between a data point and the centroid while updating the relative locations of centroids to minimize within-cluster variation.

We categorize the colors of the pixels composing a grid cell using the K-means clustering algorithm and a tuple of R, G, B values as an anonymous data point. By setting the number of clusters to *k*, a grid cell is represented by only *k* colors. The purpose of this clustering is to simplify the grid cells to be easily used as an input for the standardization process later while minimizing information loss. A grid cell consists of approximately $1200^2$ pixels with unique R, G, B values. Aggregating the colors of every pixel at once without any simplification would yield a list of $1200^2$ * 4,452 unique R, G, B values, thereby severely increasing computational cost. Simplification using K-means clustering in this step compresses the colors to the fixed number *k*, reducing the length of the list to *k* * 4,452.

The choice of *k* (the number of clusters) involves a trade-off between informational richness and computational convenience. A larger *k* depicts a more detailed picture, but reduces the benefit of color simplification and increases computational cost. On the other hand, a smaller *k* might ease the computational burden but distort the information contained in the original images. We set the number of clusters to 12 so that each image could be expressed as a combination of only 12 colors.

**Standardization.** We aggregate 53,424 colors from the previous step and apply K-means clustering once again. This process aims to create *k* standardized colors to enable comparison across all grid cells. Unlike the previous stage, the choice of *k* is contingent on how well standardized colors predict population density. We assign *k* values from 6 to 24, and the results reported are based on *k* = 12. This approach is similar to Huang et al. (2021), in which they assessed the housing quality based on categorized roof colors by K-means clustering. We extract two metrics and use them as attributes to predict the population of a grid cell: the number of pixels associated with each standardized color (*color counts*) and the Herfindahl-Hirschman index (HHI). The former captures the composition of colors, reflecting symbols on a grid-level map (e.g., roads, meadows, buildings). The latter is calculated as the squared sum of color counts and represents how concentrated or diversified the colors are within a single grid cell, capturing the complexity of an image.

**Training the models.** We train two machine learning models to predict the grid-level population. We randomly choose 20% (*n* =890) of the 4,452 grid cells as training sets. Our first approach involves using XGBoost, a state-of-the-art (regression?) model that takes tabular data as input. Color counts and the HHI values are used as input data. The model assumes a non-linear relationship between those inputs and the output but does not specify an explicit functional form for the relationship. In a second approach, we train a simple convolutional neural network (CNN) model on the same training set. CNN is one of the most widely-used machine learning models to solve image classification problems which takes an image itself as input.

Our algorithm also demonstrates the importance of the feature engineering aspect of machine learning. Feature engineering is the process of using domain knowledge about the input data and applying non-learned transformations before inputs are added to the model. Contrary to the popular conception that model development is the most important part of machine learning, experts emphasize the importance of careful pre-processing and feature engineering to prepare for high-quality input data. Simpler and more structured inputs can have a stronger impact on improving a machine learning model's prediction accuracy than tweaking and tuning the model itself[5]. Through careful feature engineering, we are able to extract information from maps without applying object detection and image segmentation algorithms.

Another advantage of our method is flexibility in data structure of target or ground truth. A typical machine learning application using map images as input requires a label for each image. As a result, grid-level ground truth data are most suitable. Our algorithm, on the other hand, extracts meaningful features from each image and they can be easily converted into regional-level inputs so that regional-level ground truth data can be employed for predictions. Most historical and contemporary data sets are available in regional-level formats, not in grid-level ones.

2.2 South Korean Data

The main data source is a collection of full-color topographic maps of South Korea produced in 2015 by NGII. The maps, which cover the entire nation, contain a wide range of information, including various human-built structures (e.g., buildings and road networks) as well as natural features (e.g., mountains and rivers) at a 1:50,000 scale. We georeference a total of 211 maps using the WGS84 (EPSG:4326)[6] coordinate system. We also georeference 214 maps in the 1970s, which are also provided by the NGII.

We further use historical maps a century ago to verify the predictive power of our approach. We use the official 1:50,000 historical maps produced by the Japanese colonial government in Korea (*Chosun omanbunil chihyungdo*) in 1918. These maps were geo-referenced and clipped to the size of 5 km by 5 km grid cells.

---

[5] As an analogy, consider making an algorithm that can learn to tell the time based on images of an analog clock (Chollet, 2021). Simply using the raw images as input would be a much more computationally expensive task. However, converting each input image into two numbers representing the polar coordinates of the clock hands greatly simplifies the complexity of learning. This is the advantage of feature engineering.
[6] Jeju Island is excluded. The map for the Danyang region (code : 36802) was not available from the original data source.

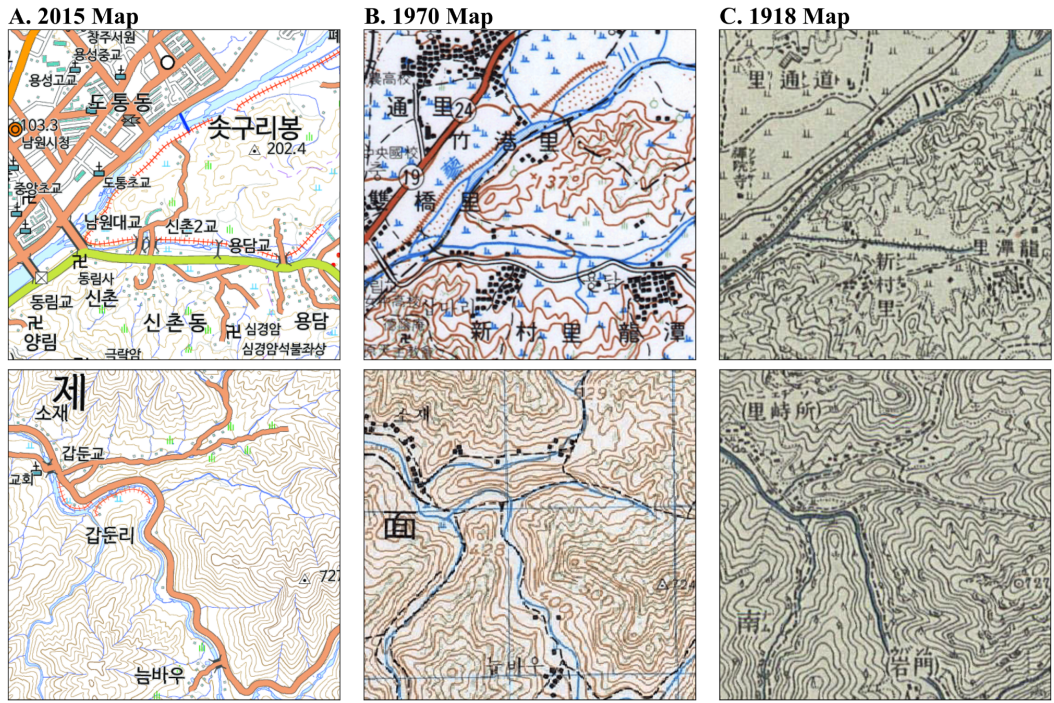

Figure 4. Samples of maps from the identical location in South Korea in year 2015, 1970, and 1918

The NGII also has provided georeferenced grid-level population data annually since 2014. We utilize a 5km by 5km grid with 2016 population data. Table 1 shows the descriptive statistics of the grid-level population data. The population density per 25 km² has a mean of 10,477 and a standard deviation of 45,215. NGII assigns zero value to grid cells with less than 6 residents due to the confidentiality problem. The distribution of the population per grid cell is skewed toward zero, with 455(9%) grid cells having zero population. This level of skewness can largely interfere with a model's learning; thus, we use the log of population as a prediction target. When only inhabited grid cells are included, the population density has a mean of 11,546 with a standard deviation of 47,336.

It is certainly impossible to get grid-level information on population density in the 1970s and 1918. Instead, we estimate the population density of each cell based on town-level census population of 1975 and 1930, under the assumption that population is uniformly distributed within a town. To validate this estimation approach, we construct the grid-level population using town-level information of 2016 in the same way and compare the estimated value with the actual grid-level population density. We find that the correlation is high, 0.75.

**Table 1:** Descriptive Statistics

|  | Obs. | Mean | Median | S.D. | Min | Max |
|---|---|---|---|---|---|---|
| Population | 4,914 | 10,477.24 | 887 | 45,215.52 | 0 | 721,105 |
| Population (inhabited grids only) | 4,459 | 11,546.35 | 1,047 | 47,336.67 | 6 | 721,105 |

Besides population density, we also utilize various outcomes as the predictands at a 5 km by 5 km grid level.

- Income and consumption: Korean Credit Bureau (KCB) has provided regional information on income and consumption at a county level since 2018. We assume that income and consumption are uniformly distributed within a county, and split those variables into 5 km grid cells. We use income data in 2018 and consumption in 2019, which are the closest time periods available to contemporary maps in 2015.

- Manufacturing employment and the number of establishments: The South Korean government provides the geographical coordinates of the population of establishments located in South Korea from 2008. We match location information with Establishment Census, which contains detailed information on establishments, such as industry code, birth year, and employment. By matching the coordinate information, we were able to identify which grid cell that establishments fall into, and count the number of establishments. We also aggregate the number of employers and use these two variables as proxies for the production level.

- Electric consumption : Korea Electric Power Corporation (KEPCO) is the sole electricity supplier in South Korea run by the government. They provide detailed information on electric consumption, including usage and industrial type at village level. The size of villages are usually smaller than 5 km grid cells, so we aggregate village-level total electric consumption to 5 km grid cells.

## 2.3 Sub-Saharan African Data

The U.S. Army Map Service has produced full-colored topographic maps in the 1940s to 1960s. We use the 1950s' maps which partially cover Sub-Saharan Africa to predict the spatial distribution of population at that time at a granular level[7]. Among 267 maps that cover

---

[7] The U.S. Army Maps are freely downloadable from University of Texas Libraries. (https://maps.lib.utexas.edu/maps/ams/)

South, West, and East Africa in a scale of 1:250,000, we georeference 75 maps of Senegal, Ghana, Guinea, and Ivory Coast[8]. The sample image of the U.S. Army Maps is on the Appendix.

We construct a training set of 1950s Sub-Saharan Africa from two sources. First, we utilize grid-level population data provided by Jedwab and Storeygard (2021). Jedwab and Storeygard (2021) obtained the exact location and decadal city-level population estimates in 33 countries in Africa since 1960. By pinning down the location of each city, they were able to create a 0.1 by 0.1 degree (approximately 11km by 11km) grid-level city population of 1960. We use these grid cells to clip georeferenced maps and match the population information with each map. Consequently, we can get 383 grid cells with the population information. Second, we utilize land cover and land use data provided by the United States Geological Survey (USGS)[9]. They provide detailed global land cover characteristics with a resolution of 1 km per pixel. We define a grid as an uninhabited area if more than 70% is covered with water bodies or more than 80% is savanna. By doing so, we can get 877 grid cells with zero population. Yet, as a distribution skewed toward zero could interfere with an effective learning of the model, we randomly select 50% of land use-based uninhabited grid cells and use them to train the model, together with the city-level population. Eventually we use 763 (7%) grid cells as a training set among 9,886 available grid maps. Figure A.2. shows the geographical distribution of the training set.

---

[8] 18 maps in Senegal, 22 in Ivory Coast, 20 in Guinea, and 15 in Ghana.
[9] https://www.usgs.gov/media/images/africa-land-cover-characteristics-data-base-version-20

## 3. Results

In this section, we compare the performance of the two models and evaluate how well they predict various economic activities at the grid level based merely on the color composition and image complexity of map data. To compare the predictive power of the two models in a consistent way, we regress the predicted population on the actual value and report $R^2$.

3.1 South Korea

3.1.1 Contemporary map in 2015

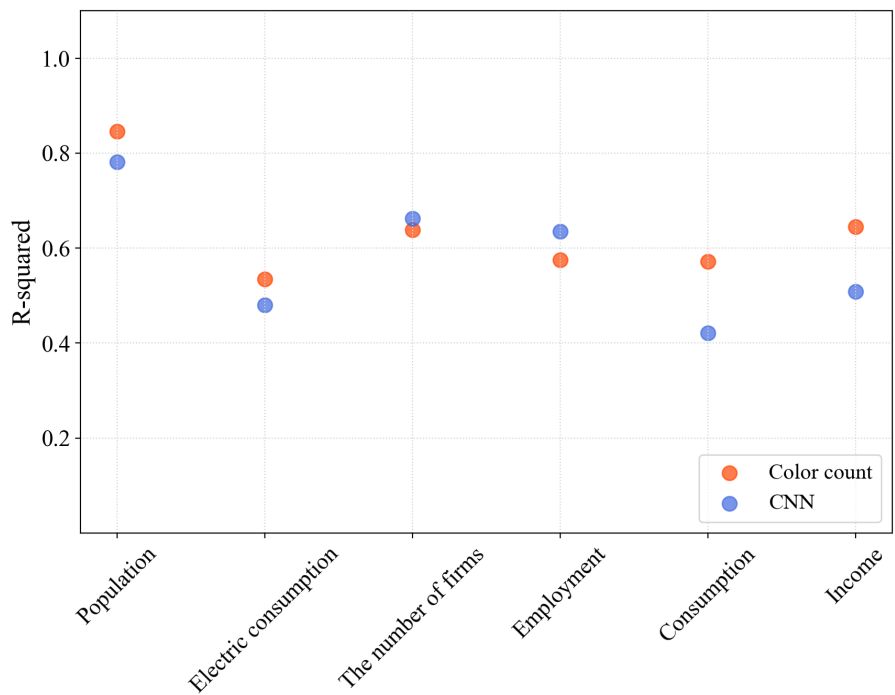

Figure 5. Comparing predictive power of our model across economic variables

Figure 5 shows the predictive performance of our models trained on 20% randomized samples when $k = 12$. We find that the color counts and HHI values are strongly predictive of the log population at the grid level. With a 20% training set, $R^2$ is 0.846 for the XGBoost model which uses color compositions and HHI as inputs, and it surpasses CNN's performance of $R^2 = 0.781$. This result indicates that the predictions based on color counts and HHI values can capture approximately 85% of the actual population's variation. (See Figure A.3.)

Although population density is a basic and traditional measure of economic activity, it is possible that maps are not suitable for predicting alternative economic variables, including production, income, or consumption level. To address this concern, we utilize a variety of data sources that can possibly measure economic activities. We find that the predictive power of our approach using color composition and image complexity is reliably stable across various economic outcomes. $R^2$ is above 0.5 consistently which is comparable to CNN or even outperforms it.

.

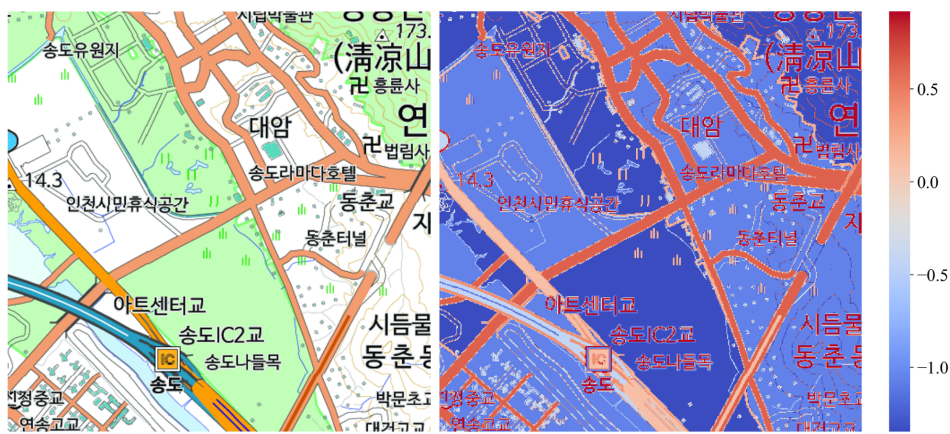

Figure 6: Interpretability of predictions

We now turn to the next exercise: disentangling the relationship between the standardized colors and the population density. We aim to understand the extent to which standardized colors play a role in predicting the population.

We estimate the following equation using the predicted log population from the linear regression model with a *k* value of 12 as a dependent variable :

$$(1) \quad Pop_i = \sum_k \alpha_k n_{k,i} + \sum_k \beta_k n_{k,i}^2 + \gamma HHI_i + \varepsilon_i$$

where $Pop_i$ is the log population of grid *i*, $n_{k,i}$ is the number of pixels in grid cell *i* classified as a standard color *k*, and $HHI_i$ is the Herfindahl-Hirschman index of a grid cell *i*. We combine the two coefficients, $\alpha_k$ and $\beta_k$, to capture how each color does play a role in the population prediction. Figure 6 is the heatmap that visualizes the relationships between standardized colors and the predicted population, along with the original map. The combined regression coefficients are distributed along with the color lamp. The stark distinction in colors shows that human-built structures (e.g. roads, buildings, and paddies) drawn in red and natural features (e.g. mountains and rivers) drawn in blue work in the opposite direction in the population prediction. Artificial structures are positively correlated with the population density whereas natural features are negatively correlated. Because different colors tend to be used for different symbols, this segmentation suggests that our algorithm can mimic object detection methods by capturing objects based on their colors.

3.1.2 Historical map in 1918

Figure 7 compares the distribution of economic activity measured by population density a century ago. Panel C is a figure of population distribution estimated from the census in 1930 which we have used as a label, and Panel D shows predicted values. The distributions are quite similar and plausible ; most people have lived in fecund *Jeonnam* and *Gyeongnam* provinces a century ago, while mountainous areas in the East show relatively lower population density.

The lower performance rate compared to contemporary maps can be due to several reasons. First, historical maps are paper-scanned and of lower quality. Second, the 1918 maps have far fewer colors compared to the modern maps. Third, we generate the grid-level population based on the assumption of uniform distribution, causing measurement errors.

A. True population density (2015)

B. Predicted population density using color counts (2015)

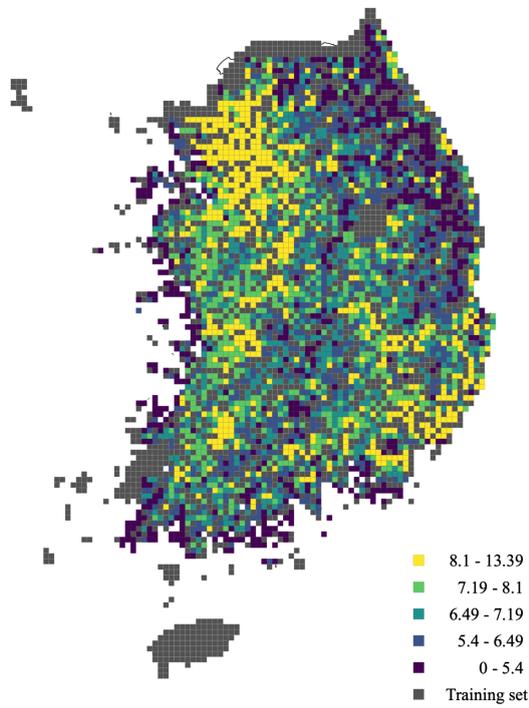
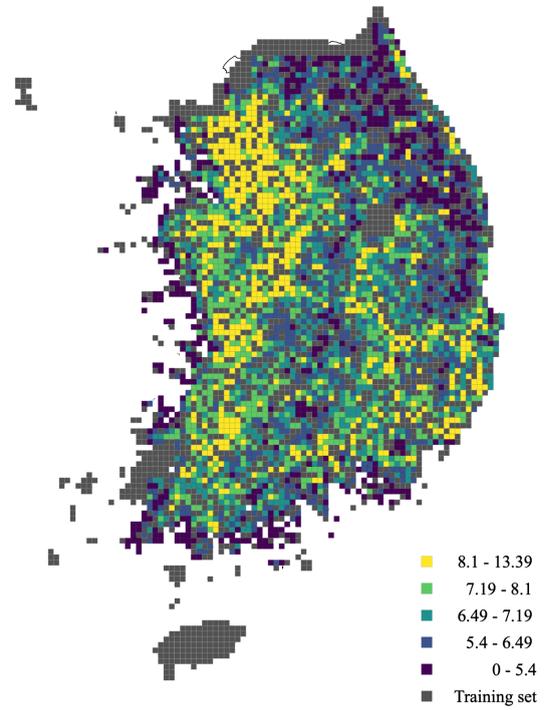

C. True population density (1918)

D. Predicted population density using color counts (1918)

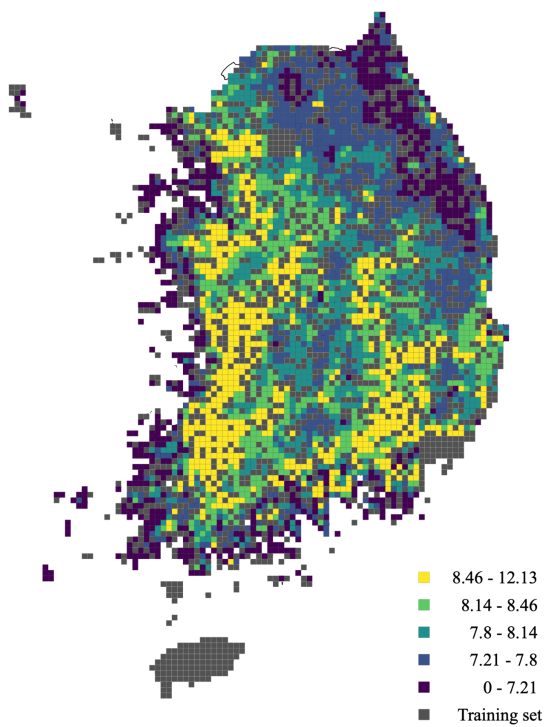
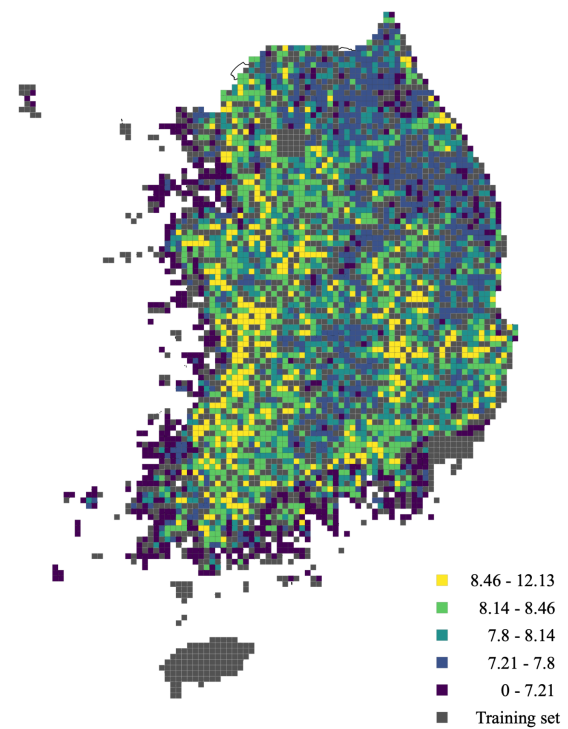

Figure 7: Spatial distribution of actual and predicted grid-level population in 2015, 1930

### 3.1.3 Predicting population growth over a century

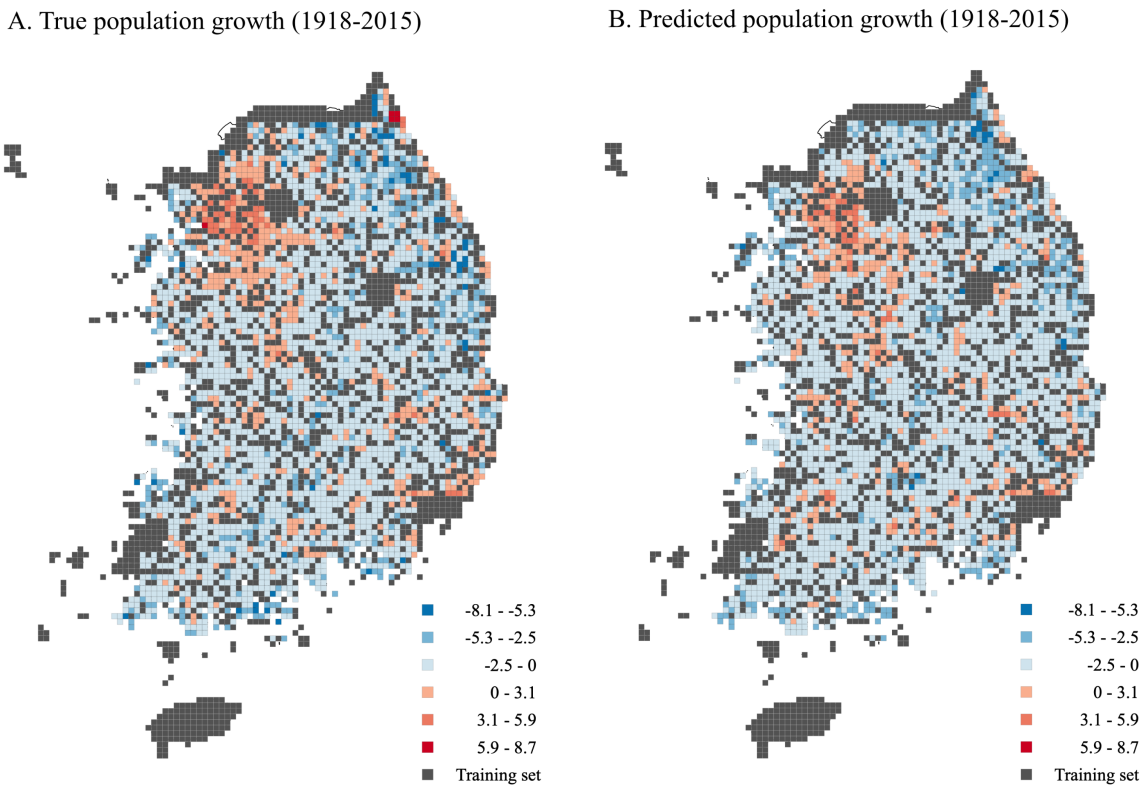

Figure 8. Spatial distribution of the population growth from 1918 to 2015

Figure 8A shows the actual distribution of log population change from 1918 to 2015, and 8A depicts the predicted one. We use the log change in population as a label and input color counts from both 1918 and 2015 to incorporate information from two periods apart. We find that our model using color counts and image complexity, which has $R^2$ of 0.639, is also predictive of the change over a century. We also attached two grid maps from 2015 and 1918 representing the same region side by side(see Figure 4A and 4C). By doing so, we were able to train CNN model and it shows a slightly lower predictive accuracy ($R^2$ = 0.608, see Figure A5 and A7).

## 3.2 Sub-Saharan Africa

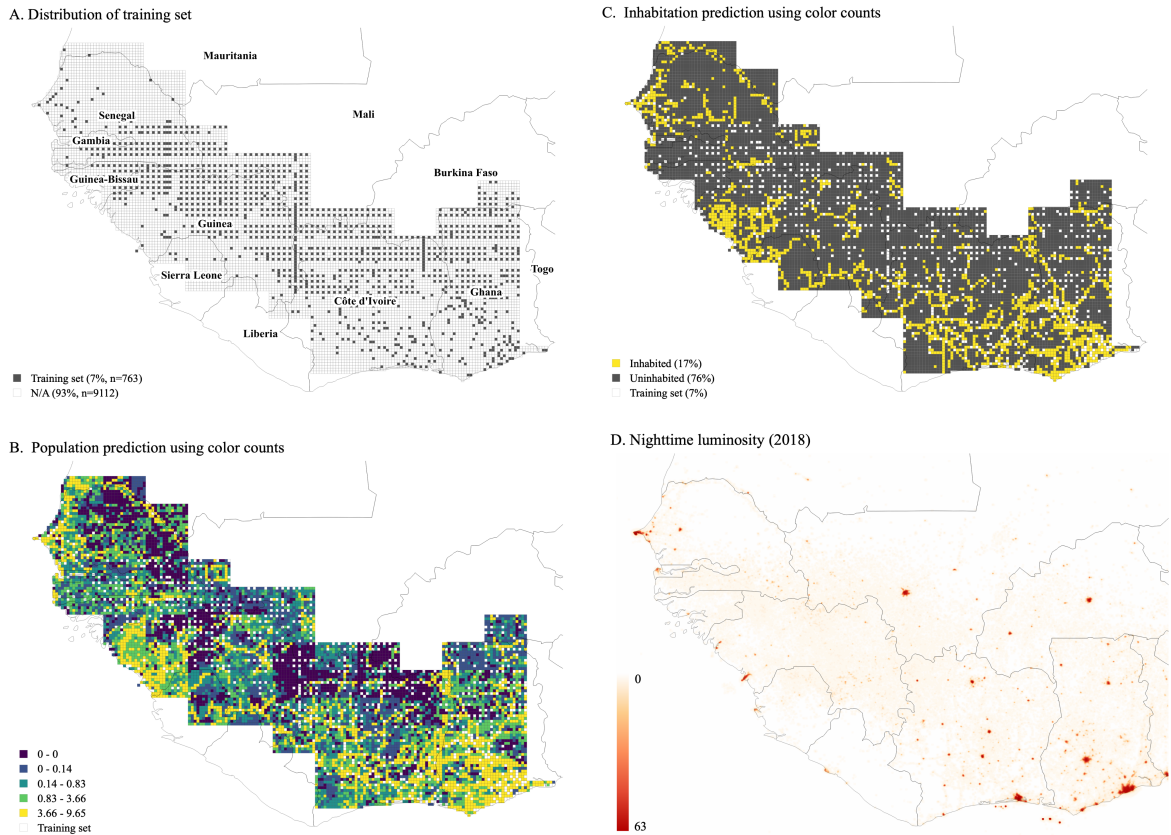

Figure 9. Prediction results of economic activity of four African countries in mid-20th century

Can we apply our model to systematically estimate the local economic activity of developing countries in the past? To answer this question, we past-cast the distribution of population density of four countries in Sub-Saharan Africa in the mid-20th century: Ghana, Guinea, Ivory Coast, and Senegal. We train our model using 80% (n=610) training set and use 20% (n=152) validation set. As we don't have information on the actual grid-level population of Sub-Saharan African countries except for the training set, we report $R^2$ when we regress the predicted population on the actual value only for validation sets. $R^2$ is 0.428 for XGBoost and 0.629 for CNN.

We also train our model to classify uninhabited and inhabited areas using identical training and validation set. We define an area as uninhabited if no one resides (zero population) and define it as inhabited otherwise. We find that both of our models predict inhabitation with high accuracy: color compositions and HHI predict 88% of uninhabited regions and 79% of inhabited ones correctly, while CNN predicts 94% of uninhabited areas and 74% of inhabited areas.

Figure 9 contains four maps that depict our models' predictive power. Panel A shows the distribution of the training set, Panel B and C are the log population and inhabited areas predicted by our model based on color counts, respectively. We replace 2,009 slightly less-than-zero predictions of population density to zero in Panel B. Panel D shows the nighttime luminosity[10] of today, a proxy widely used to measure local economic vibrancy. We found that the spatial distribution of four countries a half-century ago is correlated with today's economic vibrancy. Yet, as we can see in Panel D, the nighttime luminosity is sparsely distributed compared to the population density predicted using historical maps. This is parallel to a limitation of night light intensity data that it rarely captures economic activity below the poverty line, which is prevalent in African countries (Jean et al., 2016).

## 4. Conclusions

In this paper, we show that maps can be used to predict socioeconomic measures consistently. We solve the technical problem of using maps by extracting meaningful features from the complex combination of symbols, texts, and lines contained in a typical map. Our model can be applied in several ways. One application is to leverage maps with extensive spatial and temporal coverage, such as those produced by the U.S. Army Map Service[11] (Figure 1) across nations. In principle, our model can make a prediction on every image once they share similar color compositions with a training set. If our model is trained on a country using historically archived economic variables as labels, then it can systematically make a prediction at a granular level, even in countries lacking available historical data.

Second, our model can be applied to predict region-level economic statistics, rather than at a grid-level. Unlike existing machine learning models which require a particular type or form of inputs (e.g. CNN models use grid-level images as inputs), our model's flexibility enables us to use regional statistics as labels. This aspect is particularly important to social scientists, considering most historical data sources provide information at a regional level, such as municipalities, provinces, and counties.

---

[10] We utilize the harmonized version of nighttime luminosity data of 2018 by Li et al. (2020).
[11] https://www.loc.gov/maps/?fa=contributor:united+states.+army+map+service

A few limitations still remain. Georeferencing maps is still a labor intensive and costly prerequisite for economists to apply our model. Limited number of colors used in historical maps makes predictions noisier, particularly older maps decades ago. Yet, some possible refinements are within reach. By incorporating nation or region-wise characteristics into the model, which resembles the concept of regional fixed effects, could increase predictive power. (Park et al., 2022)

# Appendix

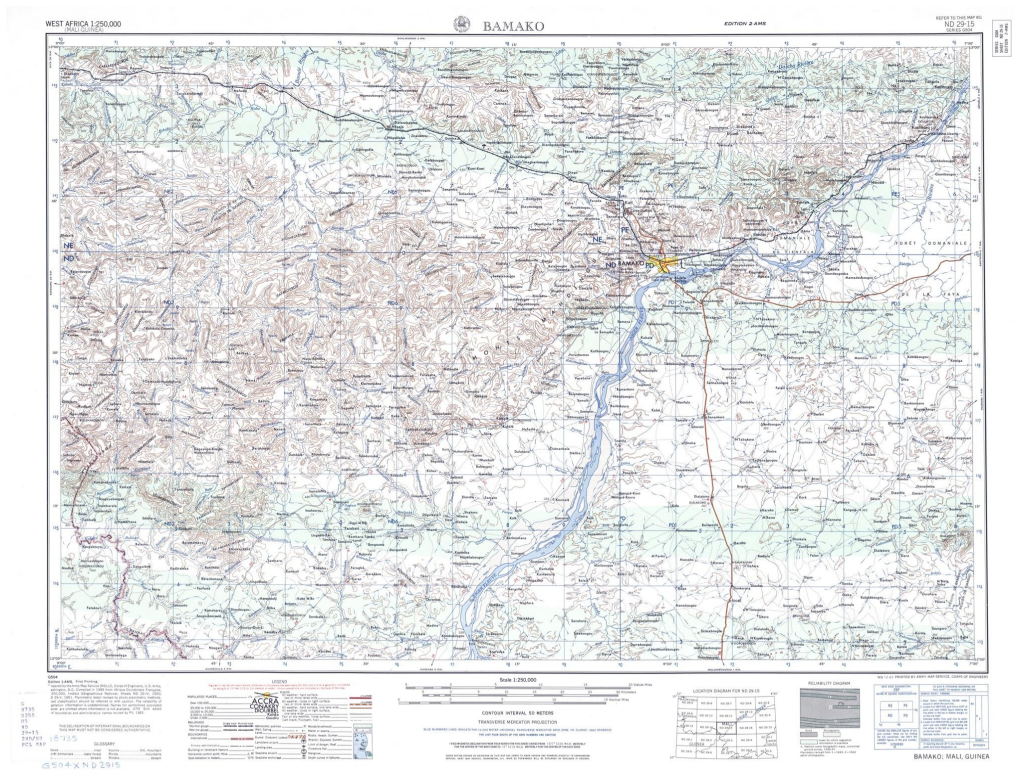

Figure A.1. Example of the U.S. Army Map

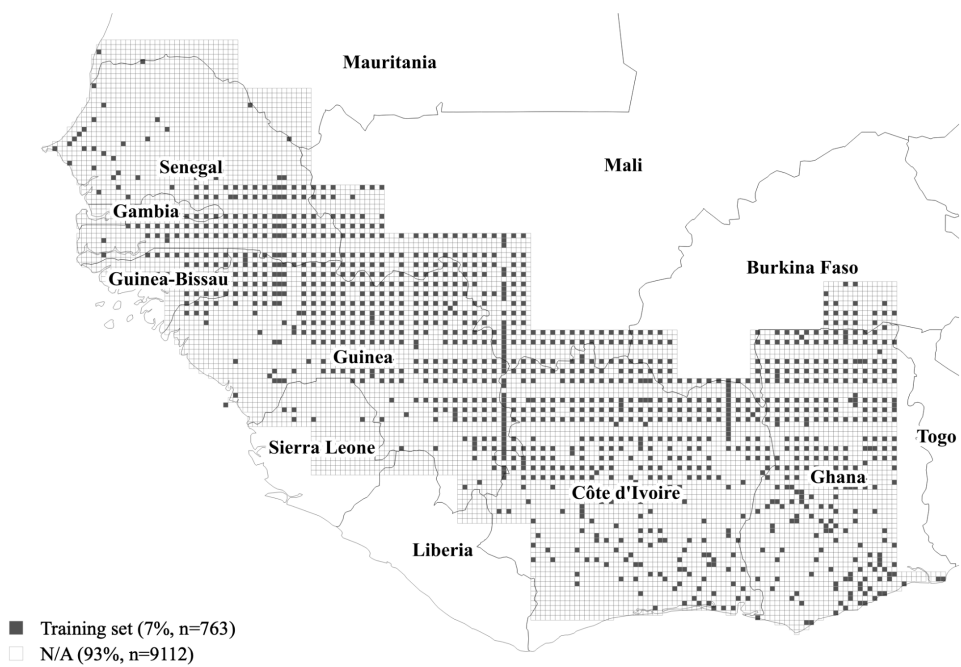

Figure A.2. Spatial distribution of training set

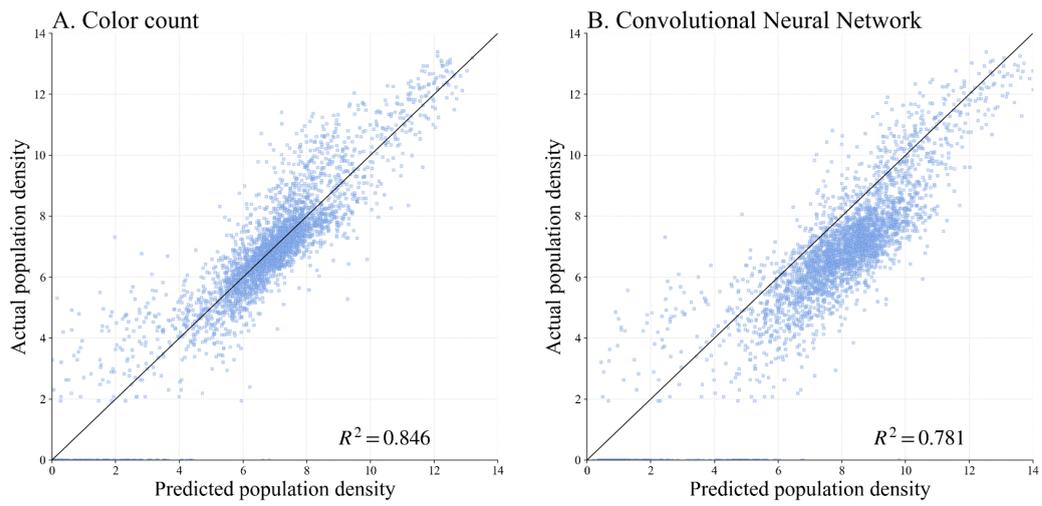

Figure A.3. Scatterplot 2015 population

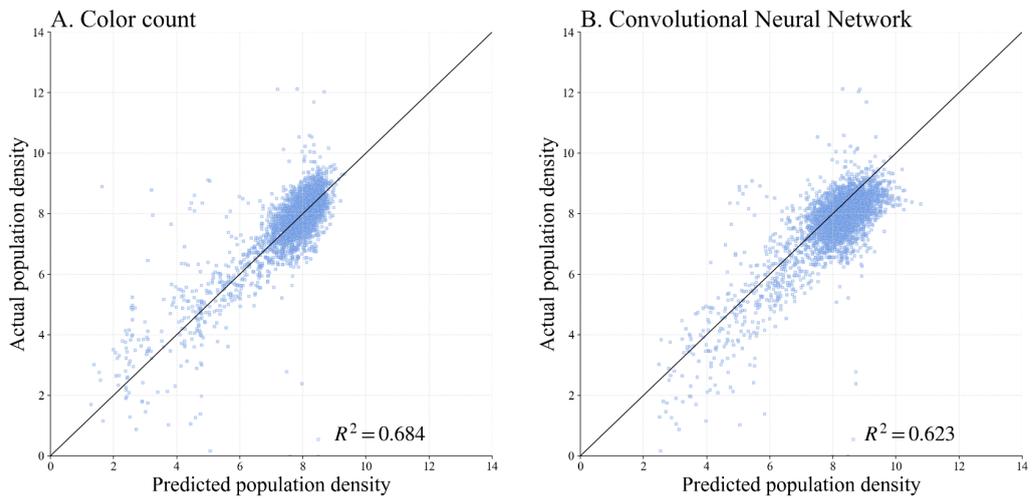

Figure A.4. Scatterplot 1918 population

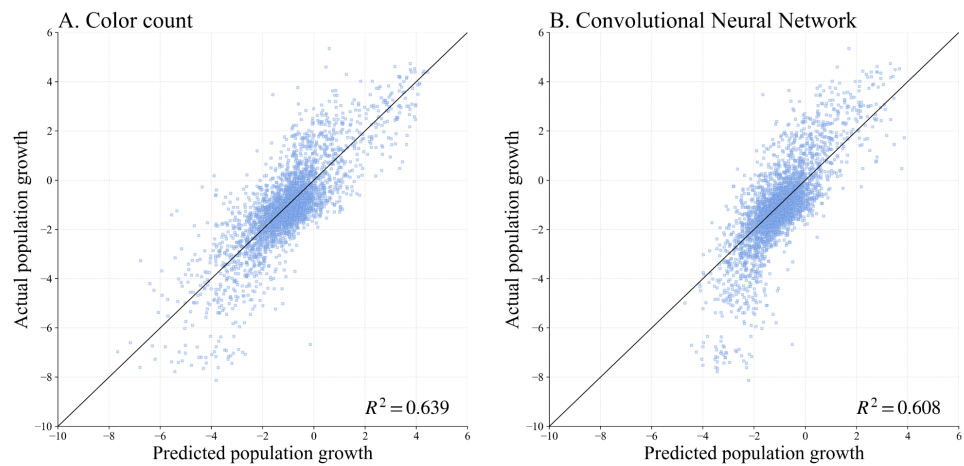

Figure A.5. Scatterplot 1918-2015 population growth

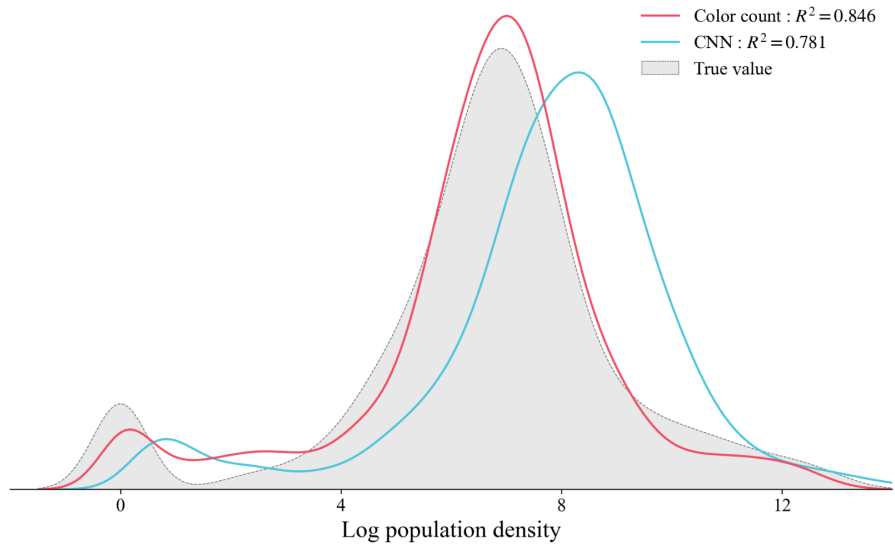

Figure A.6. Kernel density plots of grid-level actual and predicted population in 2015

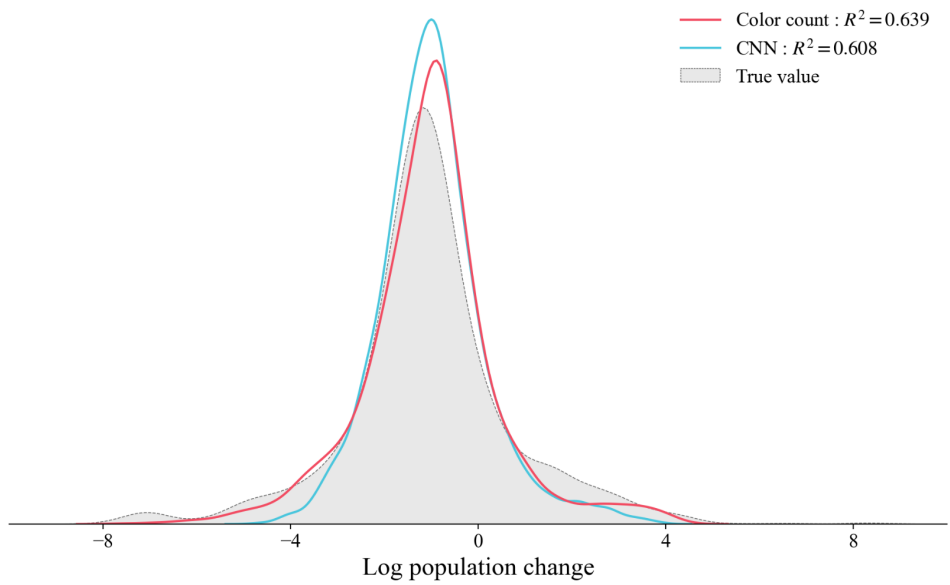

Figure A.7. Kernel density plots of grid-level actual and predicted population growth from 1918 to 2015